\documentclass[apj]{emulateapj}
\usepackage{times}
\usepackage[normalem]{ulem}
\usepackage{epsfig}
\usepackage{natbib}
\usepackage{amsfonts}
\usepackage{amsmath}
\usepackage{multirow}
\usepackage{enumerate}
\usepackage{verbatim}
\usepackage[usenames]{color}
\usepackage[plainpages=false, colorlinks=true, anchorcolor=blue, linkcolor=blue, citecolor=blue, bookmarks=false]{hyperref}
\newcommand{\be}{\begin{eqnarray}}
\newcommand{\ee}{\end{eqnarray}}
\newcommand{\lp}{\left(}
\newcommand{\rp}{\right)}

\newcommand{\slugcom}{Accepted for publication in The Astrophysical Journal}
\slugcomment{\slugcom}

\begin{document}

\normalsize


\title{Numerically Modeling the First Peak of the Type IIb SN 2016gkg}

\author{Anthony L. Piro\altaffilmark{1}}
\author{Marc Muhleisen\altaffilmark{2}}
\author{Iair Arcavi\altaffilmark{3,4,*}}
\author{David J. Sand\altaffilmark{5}}
\author{Leonardo Tartaglia\altaffilmark{5,6}}
\author{Stefano Valenti\altaffilmark{6}}

\altaffiltext{1}{The Observatories of the Carnegie Institution for Science, 813 Santa Barbara St., Pasadena, CA 91101, USA; piro@obs.carnegiescience.edu}

\altaffiltext{2}{California Institute of Technology, 1200 E California Blvd., Pasadena, CA 91125, USA}

\altaffiltext{3}{Department of Physics, University of California, Santa Barbara, CA 93106, USA}

\altaffiltext{4}{Las Cumbres Observatory Global Telescope, 6740 Cortona Dr Ste 102, Goleta, CA 93117, USA}

\altaffiltext{*}{Einstein Fellow}

\altaffiltext{5}{Texas Tech University, Physics \& Astronomy Department, Box 41051, Lubbock, TX 79409-1051, USA}

\altaffiltext{6}{Department of Physics, University of California, Davis, CA 95616, USA}

\begin{abstract}
Many Type IIb supernovae (SNe) show a prominent additional early peak in their light curves, which is generally thought to be due to the shock cooling of extended  hydrogen-rich material surrounding the helium core of the exploding star. The recent SN 2016gkg was a nearby Type IIb SN discovered shortly after explosion, which makes it an excellent candidate for studying this first peak. We numerically explode a large grid of extended envelope models and compare these to SN 2016gkg to investigate what constraints can be derived from its light curve. This includes exploring density profiles for both a convective envelope and an optically thick steady-state wind, the latter of which has not typically been considered for Type~IIb~SNe models. We find that roughly $\sim0.02\,M_\odot$ of extended material with a radius of $\approx180-260\,R_\odot$ reproduces the photometric light curve data, consistent with pre-explosion imaging. These values are independent of the assumed density profile of this material, although a convective profile provides a somewhat better fit. We infer from our modeling that the explosion must have occurred within $\approx2-3\,{\rm hrs}$ of the first observed data point, demonstrating that this event was caught very close to the moment of explosion. Nevertheless, our best-fitting one-dimensional models overpredict the earliest velocity measurements, which suggests that the hydrogen-rich material is not distributed in a spherically symmetric manner. {We compare this to the asymmetries seen in the \mbox{SN IIb} remnant Cas A, and we discuss the implications of this for Type IIb SN progenitors and explosion models.}
\end{abstract}

\keywords{
	hydrodynamics ---
	radiative transfer ---
	supernovae: general ---
	supernovae: individual: SN 2016gkg }

\section{Introduction}
Observations of supernovae (SNe) during the first few days provide valuable information about their progenitors and the circumstellar environment of the explosion \citep[][and references therein]{Piro13}. Although historically it has been difficult to catch SNe at such early moments, current and forthcoming wide-field surveys have increased the focus on early light curves. One of the exciting results from such observations is the discovery of a subclass of SNe IIb, SNe showing evidence for both hydrogen and helium in early spectroscopic observations \citep{Filippenko88,Filippenko97}, that show a ``double-peaked'' light curve, where the first peak lasts for up to a few days and the second peak lasts a couple weeks. Well-observed examples of double-peaked SNe IIb include SN 1993J \citep{Wheeler93,Richmond94}, 2011dh \citep{Arcavi11,Ergon14}, 2011fu \citep{Kumar13} and 2013df \citep{MoralesGaroffolo14,VanDyk14}. It is now generally accepted that the first peak comes from the presence of low-mass ($\sim0.01-0.1\,M_\odot$), extended (\mbox{$\sim10^{13}\,{\rm cm}$}) material \citep{Woosley94,Bersten12,Nakar14,Piro15}. This unique structure is consistent with pre-explosion imaging, which has connected the progenitors to yellow supergiants {or at least supergiants that appear much hotter than the typical red supergiants associated with hydrogen-rich Type IIP SNe \citep{Aldering94,Maund11,VanDyk14}. Such progenitors are expected for interacting binary systems \citep[e.g.,][]{Benvenuto13,Yoon17}, although see the work by \citet{Kochanek17} which argues that Cas A \citep[which is known to be a Type IIb SN from its light echoes,][]{Krause08,Rest08,Rest11,Finn16} did not have a massive binary companion at the time of explosion (a fact we revisit later in this work).}

The recent well-studied Type IIb SN 2016gkg provides an excellent opportunity to test and refine our ideas about \mbox{SNe IIb.} It was caught especially early after explosion and shows a prominent double-peaked light curve. It has well-sampled multi-band coverage including ultraviolet wavelengths, early velocity measurements of the ejecta, and pre-explosion imaging with the {\it Hubble Space Telescope} \citep{Kilpatrick17,Tartaglia17,Arcavi17}. Thus far though, most of the work analyzing its first peak has been restricted to analytic and semi-analytic models, making use of some combination of the results from \citet{Rabinak11}, \citet{Nakar14}, \citet{Piro15}, and \citet{Sapir16}. Here we extend this work by generating a large grid of models representing an extended envelope structure (in total we run 4,800 models), which are then exploded numerically for comparison with SN 2016gkg. Although the general properties we find are not qualitatively different from these previous works (we need $\sim0.02\,M_\odot$ of extended material at a radius of $\approx180-260\,R_\odot$), our calculations provide much better fits to the multi-band light curves and give some of the best constraints thus far for any SNe IIb progenitor outer structure.

In Section \ref{sec:setup}, we describe the numerical approach employed in this work. This is followed by the generation of a large grid of models, which are compared to the photometry and velocity evolution of SN 2016gkg in Section \ref{sec:compare}. We conclude in Section \ref{sec:conclusions} with a summary of our results and a discussion of the implications of our work.

\section{Explosion and Light Curve Implementation}
\label{sec:setup}

We begin by describing our methods for generating stellar models, exploding these models, and then calculating the resulting light curves. We start with a helium core that was generated from a $15\,M_\odot$ zero-age main-sequence star using the 1D stellar evolution code \texttt{MESA} \citep{Paxton13}. Using the overshooting and mixing parameters recommended by \citet{Sukhbold14}, the star is evolved until a large entropy jump between the core and envelope was established. The convective envelope is removed to mimic mass loss during a common envelope phase. The resulting helium core has a mass of $\approx4.95\,M_\odot$.

\begin{figure}
\epsscale{1.18}
\plotone{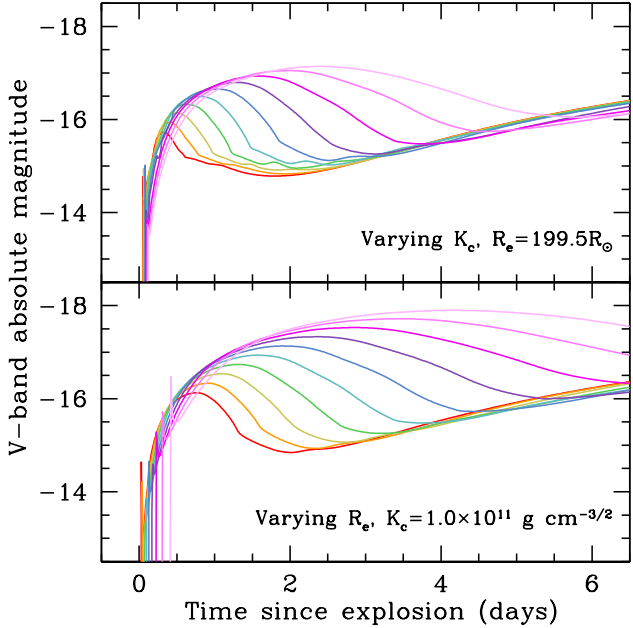}
\caption{$V$-band absolute magnitude light curves for the first peak as $K_c$ and $R_e$ are varied for a convective outer density profile. In the top panel, we fix $R_e=199.5\,{R_\odot}$ and vary $K_c=4.0\times10^9$, $6.3\times10^9$, $1.0\times10^{10}$, $1.6\times10^{10}$, $2.5\times10^{10}$, $4.0\times10^{10}$, $6.3\times10^{10}$, $1.0\times10^{11}$, $1.6\times10^{11}$. and $2.5\times10^{11}\,{\rm g\,cm^{-3/2}}$ from narrow to wide. In the bottom panel, we fix $K_c=1.0\times10^{11}\,{\rm g\,cm^{-3/2}}$ and vary $R_e=79.4$, $100.0$, $125.9$, $158.5$, $199.5$, $251.2$, $316.2$, $398.1$, $501.2$, and $631.0\,R_\odot$ from narrow to wide.}
\label{fig:photometry_multi}
\epsscale{1.0}
\end{figure}

Using this helium core, we next stitch on an extended envelope of material to mimic the hydrogen-rich material expected around SNe IIb. For this we consider both a convective envelope with density profile parameterized as
\be
	\rho_c(r) = K_c/r^{3/2}
\ee
and a steady wind profile parameterized as
\be
	\rho_w(r) = K_w/r^2.
\ee
Here the constant factor is connected to the properties of the wind by
\be
	K_w=\dot{M}/4\pi v,
	\label{eq:mdot}
\ee
where $\dot{M}$ is the mass loss rate and $v$ is the wind's velocity. In each case the density profile extends down until it connects smoothly with the underlying stellar model, and it extends out to a radius $R_e$ where it is abruptly set to zero. The composition is taken to be solar. The convective model would best represent what is typically found in binary evolution models that try to generate SNe IIb progenitors self-consistently \citep[e.g.,][]{Benvenuto13,Yoon17}. {As far as we know, a wind profile has not been considered before for SNe IIb progenitors. Our motivation for investigating it here is that when a wind is optically thick it might mimic an envelope, and it is plausible that if the SN IIb progenitor is in the midst of a mass transfer event as it explodes, then the material around the helium core might be better represented as a wind rather than an envelope.}

These models are then exploded with our open-source numerical code \texttt{SNEC} \citep{morozova:15}. We assume that the inner $1.4\,M_{\odot}$ of the models form a neutron star and excise this region before the explosion. A $^{56}$Ni mass of $0.2\,M_\odot$ is placed at the inner edge of the ejecta, with the exact value not being critical to our calculation because we only compare the first $\approx3.5\,{\rm days}$ of the calculations with the data. The compositional profiles are smoothed using a ``boxcar'' approach with the same parameters as in \citet{morozova:15}, and unlike our previous Type II calculations we do not use an opacity floor. This is important for having the correct drop of opacity as the hydrogen-rich extended material recombines. We use a ``thermal bomb mechanism'' for the explosion, where a luminosity is provided to the inner region of the ejecta to generate the explosion. We add an energy of the bomb to the internal energy in the inner $0.02\,M_{\odot}$ of the model for a duration of $1\,{\rm s}$ such that the final energy of the explosion is $E=10^{51}E_{51}\,{\rm erg}$. {In the work of \citet{morozova:15,morozova:16,Morozova17}, we explore a range of durations around this timescale and do not find a considerable differences in the light curves properties. The equation of state includes  contributions from ions, electrons and radiation, with the degeneracy\ effects taken into account as in \citet{paczynski:83}. We trace the ionization fractions of hydrogen and helium solving the Saha equations in the non-degenerate approximation as proposed in \citet{zaghloul:00}. The numerical grid consists of 1000 cells. To ensure convergence, we tested models down to a grid of 400 cells without noticeable changes in the light curves.} Photometric magnitudes are estimated from the simulations by assuming a blackbody spectrum for the emission and integrating over each of the desired wavebands (as opposed to just taking the model flux at some effective wavelength).

Figure \ref{fig:photometry_multi} highlights how the properties of the early peak change as the variable $K_c$ and $R_e$ are varied for a convective density profile. In the top panel, we vary $K_c$ and keep $R_e$ fixed. This mostly changes the width of the first peak because the mass in the convective envelope, which is given by
\be
	M_c &=& \int_{R_*}^{R_e} 4\pi r^2\rho_c(r) dr \approx \frac{8\pi}{3} K_c R_e^{3/2},
	\nonumber
	\\
	&\approx& 2\times10^{-2} K_{c,11}R_{200}^{3/2}\,M_\odot,
	\label{eq:mc}
\ee
where $K_{c,11} = K_c/10^{11}\,{\rm g\,cm^{-3/2}}$ and $R_{200} = R_e/200\,R_\odot$, sets the diffusion time of photons through the extended material. In the bottom panel, we vary $R_e$ and keep $K_c$ fixed. As $R_e$ increases the first peak becomes brighter, consistent with semi-analytic expectations \citep{Nakar14,Piro15}. The width also changes because the amount of mass in the extended material also increases as shown by Equation (\ref{eq:mc}).

In the case of a wind profile, the width and height of the first peak vary with $R_e$ and $K_w$ in a similar way to the convective profile, thus we do not provide actual light curve plots for this density profile here. We do investigate the differences between the convective and wind light curves further below. The mass of the optically thick wind has a different dependence with the extended radius, given by
\be
	M_w &=& \int_{R_*}^{R_e} 4\pi r^2\rho_w(r) dr \approx 4\pi K_w R_e,
	\nonumber
	\\
	&\approx& 8\times10^{-3} K_{w,17}R_{200}\,M_\odot,
	\label{eq:mw}
\ee
where $K_{w,17} = K_w/10^{17}\,{\rm g\,cm^{-1}}$.

\section{Comparing SN 2016gkg to Numerical Models}
\label{sec:compare}

Photometric data for SN 2016gkg was mostly taken from \citet{Arcavi17}. This work should be consulted for the full details, but to quickly summarize, the data is compiled from the discovery report by A. Buso and S. Otero, publicly-available early observations taken with the Las Cumbres Observatory \citep[LCO;][]{Brown13} global telescope network and the All-Sky Automated Survey for Supernovae \citep[ASAS-SN;][]{Shappee14}, publicly-available {\it Swift} UVOT data, the Advanced Technology Large Aperture Space Telescope \citep[ATLAS;][]{Tonry11} early-time detections, and an intensive followup campaign with LCO.

We adopt a distance of 26.4 Mpc and a distance modulus of 32.11 magnitudes to SN 2016gkg, based on Tully- Fisher distance measurements to its host galaxy NGC 613 \citep{Nasonova11}. Extinction corrections are included as the nominal value found in \citep{Tartaglia17}. The resulting multi-band photometric data is later shown in both Figures \ref{fig:photometry} and \ref{fig:photometry_wind} when we discuss our best fitting numerical models.

Of special note are the early photometric points from A. Buso, which are shown with a black circle filled in turquoise. SN2016gkg was discovered by A. Buso on Sep 20.19 UT\footnote{http://ooruri.kusastro.kyoto-u.ac.jp/mailarchive/vsnet-alert/20188} and reported by A. Buso and S. Otero\footnote{https://wis-tns.weizmann.ac.il/object/2016gkg}. The first point, estimated at 19th magnitude in a clear filter, is extremely important for constraining the models, and thus we include it in our analysis even though it is currently only reported in the included link and not published. Though the band is unfiltered, we model it as $g$-band given the early, hot phase of the SN. The exact band is not crucial for the results of our modeling. We tried other bands as well and did not see strong changes in our fits as long as the assumed band is on the Rayleigh-Jeans side of the spectrum. The key is just that this data point is included because it is important for informing us on how quickly the light curve is rising at these early phases.

\subsection{Convective Envelope Models}
\label{sec:convective}

We first consider fitting envelope models with a convective density profile to SN 2016gkg. We include 20 different radii with $R_e=12.6$, $15.8$, $20.0$, $25.1$, $31.6$, $39.8$, $50.1$, $63.1$, $79.4$, $100.0$, $125.9$, $158.5$, $199.5$, $251.2$, $316.2$, $398.1$, $501.2$, $631.0$, $794.4$, and $1000.0\,R_\odot$, and 15 different density scalings with $K_c=4.0\times10^{9}$, $6.3\times10^{9}$, $1.0\times10^{10}$, $1.6\times10^{10}$, $2.5\times10^{10}$, $4.0\times10^{10}$, $6.3\times10^{10}$, $1.0\times10^{11}$, $1.6\times10^{11}$, $2.5\times10^{11}$, $4.0\times10^{11}$, $6.3\times10^{11}$, $1.0\times10^{12}$, $1.6\times10^{12}$, and $2.5\times10^{12}\,{\rm g\,cm^{-3/2}}$. These are exploded at 8 different energies of $E_{51}=0.6$, $0.8$, $1.0$, $1.3$, $1.6$, $2.0$, $2.3$, and $2.6$. Thus in all we ran 2,400 convective envelope explosion models.

\begin{figure}
\epsscale{1.18}
\plotone{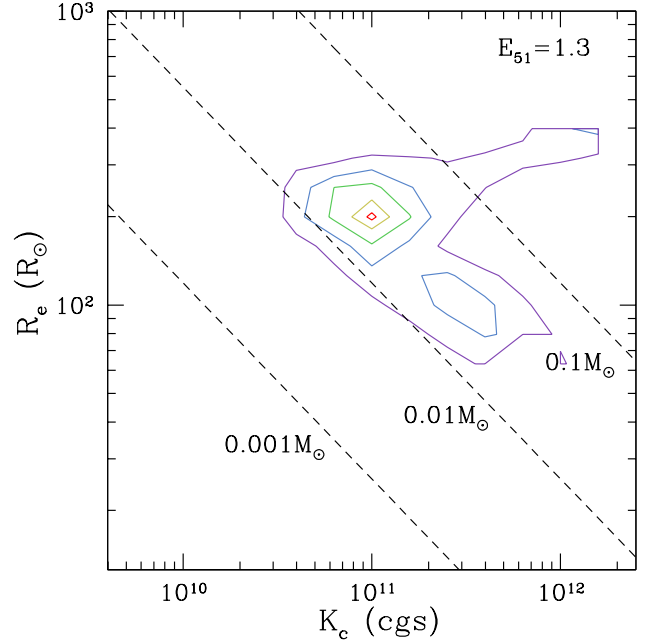}
\caption{Contours of $1\sigma$ (red), $2\sigma$ (yellow), $3\sigma$ (green), $4\sigma$ (blue), and $5\sigma$ (purple) for an explosion with $E_{51}=1.3$ and a convective density profile. The best fit considered model has $R_e=199.5\,R_\odot$ and $K_c=1.0\times10^{11}\,{\rm g\,cm^{-3/2}}$, which corresponds to $M_c\approx0.02\,M_\odot$. Lines of constant envelope mass (black, dashed lines) are drawn using Equation (\ref{eq:mc}). }
\label{fig:chisq}
\epsscale{1.0}
\end{figure}

The models were compared to the data only over the first $3.5\,{\rm days}$ so that we focus on just the first peak of the SN. In this way we are not sensitive to the details of the amount, location, or mixing of $^{56}$Ni, which would influence the rise to the second peak. This comparison was evaluated with a simple $\chi^2$ calculation, where we take
\be
	\chi^2 = \sum_i (M_{i,\rm observed} - M_{i,\rm model})^2/\Delta M_{i,\rm observed}^2,
\ee
where $M_{i,\rm observed}$ and $M_{i,\rm model}$ are the observed and model absolute magnitudes, respectively,  $\Delta M_{i,\rm observed}$ is the observed magnitude error, and the index $i$ runs over all data points in all photometric bands. In addition, because of their constraining nature, we require the model to go through the first two data points by A. Buso and S. Otero. For each calculated model, we search through various potential explosion times using a bisectional algorithm until we find an explosion time that minimizes $\chi^2$. This is taken to be the $\chi^2$ for that particular model. Once the full grid of models is run, we can identify the minimum $\chi^2_{\rm min}$ for the entire grid. The next step is how to interpret these values of $\chi^2_{\rm min}$. The issue is that the models will never be an exact fit to nature, and in addition there are systematic uncertainties in both the modeling and the data, as well as the fact that the grid spacing is not infinitely small. Therefore, we should not necessarily find $\chi^2_{\rm min}=1$. Our approach to this problem is to consider $\chi^2_{\rm min}$ as basically the best we can do and therefore effectively treat $\chi^2_{\rm min}=1$. From this we can then estimate $1\sigma$, $2\sigma$, and $3\sigma$ uncertainties as models with $\chi^2<2\chi^2_{\rm min}$, $\chi^2<5\chi^2_{\rm min}$, and $\chi^2<10\chi^2_{\rm min}$, respectively.

The results of applying this procedure are shown for the particular explosion energy $E_{51}=1.3$ in Figure \ref{fig:chisq}. There are 300 models considered across this panel. The best fit model corresponds to the red region with a value $R_e=199.5\,R_\odot$ and $K_c=1.0\times10^{11}\,{\rm g\,cm^{-3/2}}$, so that $M_c\approx0.02\,M_\odot$. We note though that given the coarseness of our grid (see the values listed above) and uncertainties in the modeling, there still remains at least a $20\%$ error in these quantities.

Also plotted in Figure \ref{fig:chisq} are lines of constant $M_c$ using Equation (\ref{eq:mc}). From this we see there are two degeneracies acting in Figure \ref{fig:chisq}. The first is at roughly constant $R_e$. This is simply set by the maximum luminosity of the first peak. The second runs along at roughly constant $M_c$, which is set by the width of the first peak. Such degeneracies are expected from the scalings described in more detail by \citet{Piro15}, but it is reassuring for our fitting routine here that they naturally appear. This lends some robustness to our derived parameters for the envelope material even if there are uncertainties in the modeling in detail.

\begin{figure}
\epsscale{1.18}
\plotone{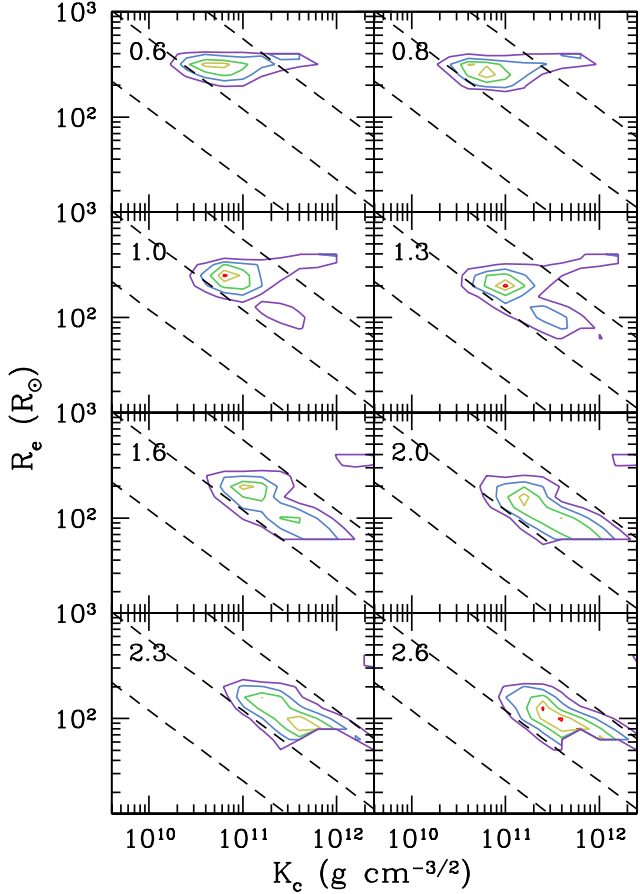}
\caption{Contours of $1\sigma$ (red), $2\sigma$ (yellow), $3\sigma$ (green), $4\sigma$ (blue), and $5\sigma$ (purple) for an explosion with $E_{51}=0.6$, $0.8$, $1.0$, $1.3$, $1.6$, $2.0$, $2.3$, and $2.6$ (as labeled) and a convective density profile. This demonstrates that there are many similarly well-fit models over a range of reasonable energies, although the absolutely best fit model is at $E_{51}=1.3$, as shown previously in \mbox{Figure \ref{fig:chisq}.} Although not labeled here, lines of constant $M_c$ run from $0.001$, $0.01$, and $0.1\,M_\odot$ from left to right in each panel (dashed, black lines).}
\label{fig:chisq_multi}
\epsscale{1.0}
\end{figure}

{In Figure \ref{fig:chisq_multi}, we consider the $\chi^2$ contours across all considered energies. This shows that although the energy we considered in Figure \ref{fig:chisq} of $E_{51}=1.3$ gives overall the best fit, an energy of $E_{51}=1.0$ can give a similarly good fit with a slight larger radius. One can also see that as the energy is varied different degeneracies gain or lose strength. At low energy, most of the degeneracy is for fixed $R_e$ because these models have difficulty matching the width of the first peak. As the energy increases, the width is better matched and a degeneracy at constant $M_c$ grows stronger. Nevertheless, independent of these issues, this demonstrates how robustly $M_c$ and $R_e$ can be constrained from this modeling with values of $\sim0.02\,M_\odot$ and $\approx180-260\,R_\odot$, respectively, where we take the spread for the value of $R_e$ as roughly our $3\sigma$ uncertainties. Distance is an additional uncertainty not included in this estimate. Since $R_e$ scales proportional to the luminosity at peak, for a distance $D$ the scaling is roughly $R_e\propto D^{2}$.}

It should be noted that there is also a degeneracy between the explosion energy and the mass of the core, because the larger the core is, the less energy there is available for the envelope. Therefore, one should really think about the energy provided by the shock to the envelope, which scales as  \citep{Nakar14}
\be
	E_e \approx2\times10^{49}E_{51}\lp\frac{M_{\rm core}}{3\,M_\odot} \rp^{-0.7}\lp \frac{M_c}{0.01\,M_\odot}\rp^{0.7}{\rm erg}.
	\label{eq:ee}
\ee
This is an important distinction because the early time evolution and the associated velocities will better reflect $E_e$ rather than $E_{51}$. For our best fit energy of $E_{51}=1.3$ and $M_{\rm core}=3.55\,M_\odot$ (once the neutron star mass is subtracted off), we estimate from Equation (\ref{eq:ee}) that $E_e\approx 3.8\times10^{49}{\rm erg}$. Thus a larger or smaller $E_{51}$ would be expected for a smaller or larger $M_{\rm core}$, respectively, to keep this value of $E_e$ roughly fixed.

\begin{figure}
\epsscale{1.18}
\plotone{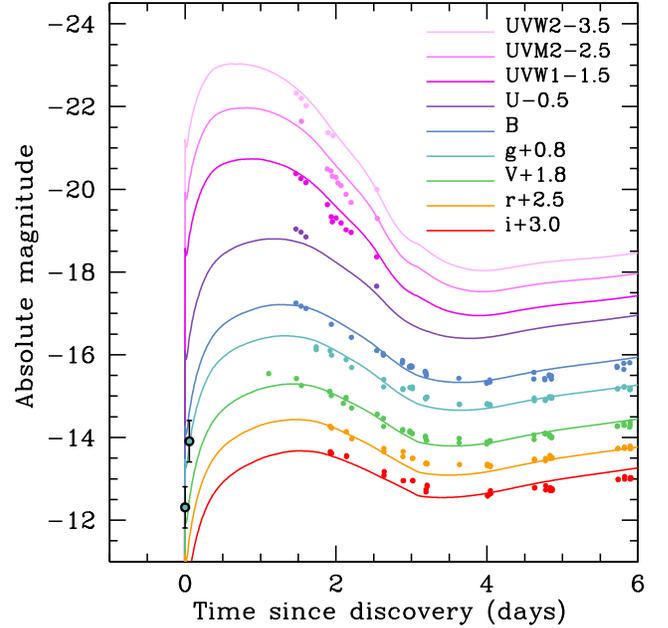}
\caption{Comparison of the multi-band data of SN 2016gkg \citep{Arcavi17} with our best fit convective envelope model using $E_{51}=1.3$, $R_e=199.5\,R_\odot$, and $K_c=1.0\times10^{11}\,{\rm g\,cm^{-3/2}}$, so that $M_c\approx0.02\,M_\odot$. Early data points from A. Buso and S. Otero are emphasized with a black outline.}
\label{fig:photometry}
\epsscale{1.0}
\end{figure}

The overall best fit model is shown in comparison to the multi-band data in Figure \ref{fig:photometry}. In comparison to previous semi-analytic fits to the data, the numerical models do a better job of following the changes from short to long wavelengths. In particular, the model by \citet{Piro15} predicts {a much more symmetrical peak,} which does an okay job of representing the data at optical wavelength, but does increasingly worse at shorter wavelengths. This is because the accelerating shock velocity in the decreasing density near the surface of the envelope causes a stronger temperature evolution than predicted in the one-zone model by \citet{Piro15}, which does not follow this velocity gradient.

Furthermore, this comparison demonstrates how crucial the early data from A. Buso and S. Otero are for constraining the rise of the first peak and thus the envelope model. {\it In fact, we infer from our modeling that the explosion must have occurred within $\approx2-3\,{\rm hrs}$ of this first data point!} This is very  close to the moment of explosion; for example, SN 2011fe was also observed very early as well and this was at roughly 4\,hrs \citep{Bloom12}. As wide field, transient surveys grow in the future, this work demonstrates the powerful constraints we will be able to provide once more early time data is available.

Another aspect to note is that the best fit extended mass of $0.02\,M_\odot$ is much less than the values around $\sim0.1\,M_\odot$ that are typically presented by \citet{Woosley94} and \citet{Bersten12} for SNe IIb. This is because the first peak is most sensitive to only the mass near the maximum radius and not the total amount of hydrogen present (see the more detailed discussion in \citealp{Nakar14}, and in particular their Figure 2, which explicitly shows how the mass measured by the first peak compares with the total hydrogen shell mass). So while the total hydrogen mass can indeed be $\sim0.1\,M_\odot$ to produce a realistic, hydrostatic model of a convective envelope, the first peak itself only can be utilized to measure the outer $0.02\,M_\odot$ of material.

\subsection{Wind Models}
\label{sec:wind}

\begin{figure}
\epsscale{1.18}
\plotone{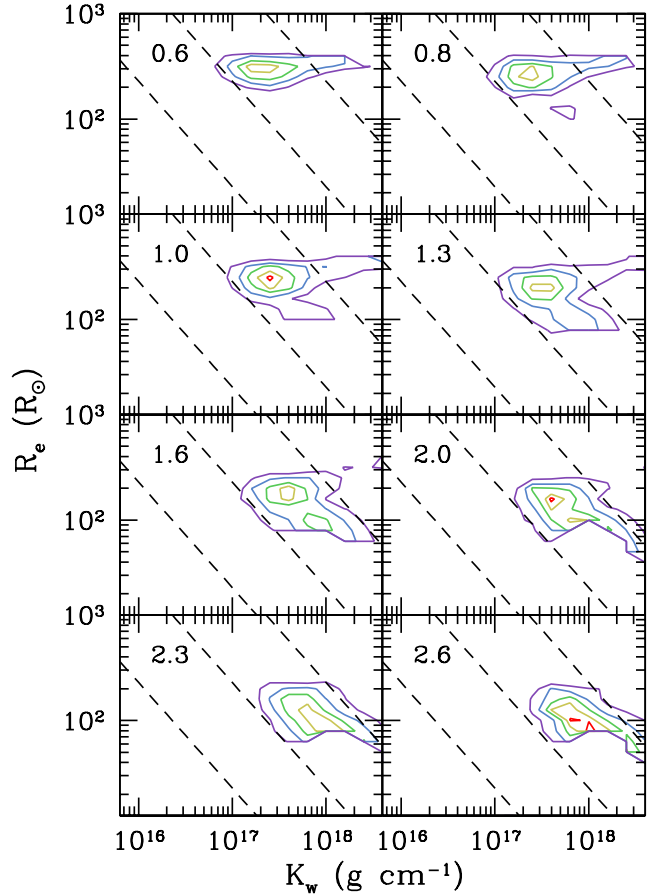}
\caption{The same as Figure \ref{fig:chisq_multi}, but now with a wind density profile. The overall best fit model is at $E_{51}=1.6$, with $R_e=199.5\,R_\odot$ and $K_w=4.0\times10^{17}\,{\rm g\,cm^{-1}}$, which corresponds to $M_w\approx0.03\,M_\odot$ using Equation (\ref{eq:mw}). Lines of constant $M_w$ run from $0.001$, $0.01$, and $0.1\,M_\odot$ from left to right in each panel (dashed, black lines).}
\label{fig:chisq_wind_multi}
\epsscale{1.0}
\end{figure}

We next consider a wind-like density profile for the envelope material. Traditionally SN IIb progenitors are thought to have extended, convective envelopes due to a recent mass transfer event that stripped the majority of its hydrogen-rich envelope \citep[e.g.,][]{Benvenuto13,Yoon17}. Nevertheless, if a star is in the midst of a mass transfer event, and the mass loss rate is great enough, the progenitor could in principle look like a yellow supergiant just from the optically thick wind. This is not traditionally considered for SNe IIb, but we explore such a density profile here in case assumptions about the density profile introduce uncertainties in the parameters estimations.

For our grid of wind models we consider the same 20 values for $R_e$ and 8 values for $E_{51}$ as for the convective models discussed in Section \ref{sec:convective}. For the mass loading factor, we again consider 15 values with $K_w=6.3\times10^{15}$, $1.0\times10^{16}$, $1.6\times10^{16}$, $2.5\times10^{16}$, $4.0\times10^{16}$, $6.3\times10^{16}$, $1.0\times10^{17}$, $1.6\times10^{17}$, $2.5\times10^{17}$, $4.0\times10^{17}$, $6.3\times10^{17}$, $1.0\times10^{18}$, $1.6\times10^{18}$, $2.5\times10^{18}$, and $4.0\times10^{18}\,{\rm g\,cm^{-1}}$. This again constitutes 2,400 different wind models.

\begin{figure}
\epsscale{1.18}
\plotone{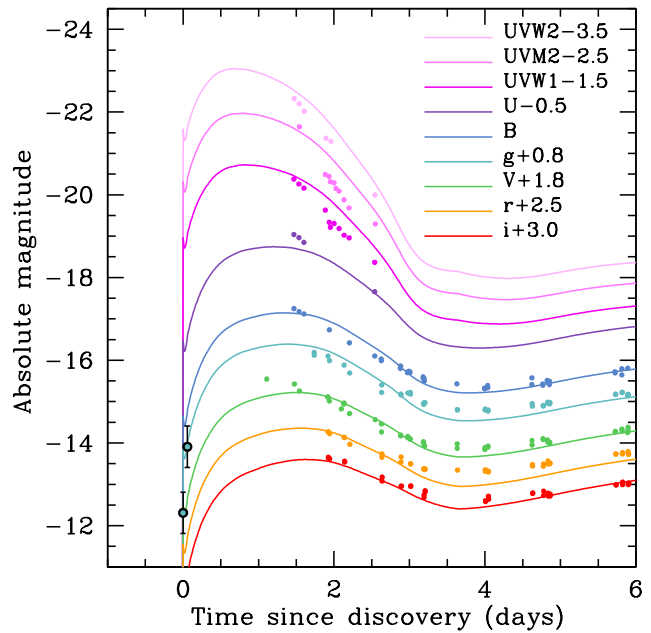}
\caption{Same as Figure \ref{fig:photometry}, but this time comparing to our best fit wind model with $E_{51}=1.0$, $R_e=251.2\,R_\odot$ and $K_w=2.5\times10^{17}\,{\rm g\,cm^{-1}}$, corresponding to $M_w\approx0.03\,M_\odot$}
\label{fig:photometry_wind}
\epsscale{1.0}
\end{figure}

The comparison of the wind models with the data is summarized in Figure \ref{fig:chisq_wind_multi}, plotted in the same way as before in Figure \ref{fig:chisq_multi}. The lines of constant $M_w$ use Equation (\ref{eq:mw}), and one can see that the degeneracy in $M_e$ (especially at larger $E_{51}$) follow the slope of these lines. Overall,  this he best fit model is at $E_{51}=1.6$, with $R_e=251.2\,R_\odot$ and $K_w=2.5\times10^{17}\,{\rm g\,cm^{-1}}$, which corresponds to $M_w\approx0.03\,M_\odot$. This is remarkably close to the result using a convective density profile, especially considering the coarseness of the grid. This demonstrates a strong constraint on these quantities, independent of the density profile.


Even though the best fit convective and wind density profile models have essentially the same radii and mass associated with them, the best fit light curve for a wind profile is compared to the data in \mbox{Figure \ref{fig:photometry_wind}.} This shows that the wind profile gives a noticeably worse fit that the convective profile. The decline from the peak is just too steep in comparison to the wind profile model, perhaps arguing that such a profile cannot explain the data. In addition, it is important to ask whether such a wind model is even physically plausible. Using Equation (\ref{eq:mdot}), and an estimated wind velocity of $v\approx100\,{\rm km\,s^{-1}}$, the corresponding mass loss rate would be $\dot{M}\approx0.8\,M_\odot\,{\rm yr^{-1}}$. This appears very high compared to what is normally expected for massive stars, although maybe it represents a star in the midst of mass transfer to a close binary companion as is expected to take place for the progenitors of SNe IIb. Furthermore, the length of time implied by the radius of the extended material is $\sim\,$weeks. With such a short timescale, it is implausible that the presence of enhanced wind generation should randomly occur so close to explosion. Either the two are casually linked, or perhaps the wind model does not really happen in nature. Future work should be done to better understand whether a wind environment as inferred here could actually be present in these binary scenarios.

\subsection{Photospheric Velocities}

Besides having early photometry, SN 2016gkg was also exceptional because it had especially early spectra and velocity measurements in comparison to other SNe IIb. This provides a unique opportunity to use early velocities as another constraint on these models.

\begin{figure}
\epsscale{1.18}
\plotone{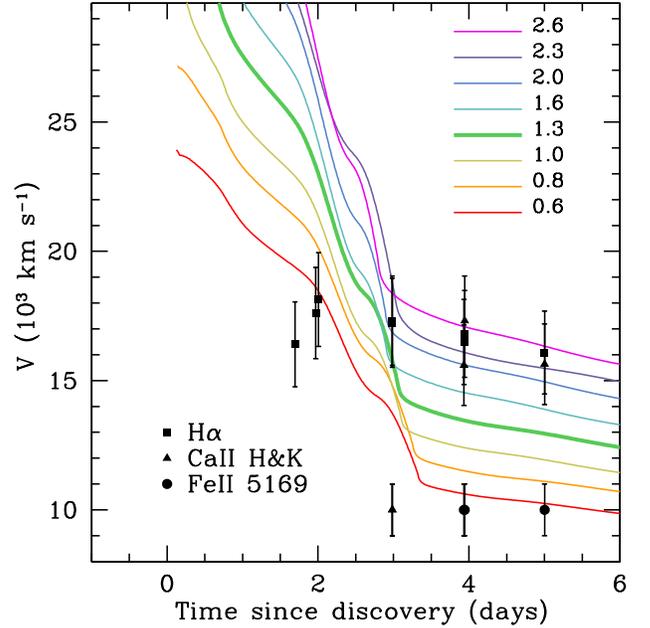}
\caption{Velocity data from \citet{Tartaglia17} for a representative range of spectral lines (as labeled) in comparison to the best fit convective density model for each considered energy. The best fit model with $E_{51}=1.3$ is highlighted with a thicker line.}
\label{fig:velocities}
\epsscale{1.0}
\end{figure}

In Figure \ref{fig:velocities}, we plot velocity data for SN 2016gkg from \citet{Tartaglia17}, which  were computed by measuring the positions of the minima of the P-Cygni absorption components. This is not all the lines that were measured, rather this is meant to present a representation of the fastest (${\rm H}\alpha$) and slowest (${\rm CaII\,H\&K}$ and ${\rm FeII\,5169}$) features observed during these early phases. Nominally, one would expect that these velocities represent an upper limit to the photospheric velocity, where the photospheric velocity roughly corresponds to where the continuum emission is mostly being generated.

\begin{figure}
\epsscale{1.18}
\plotone{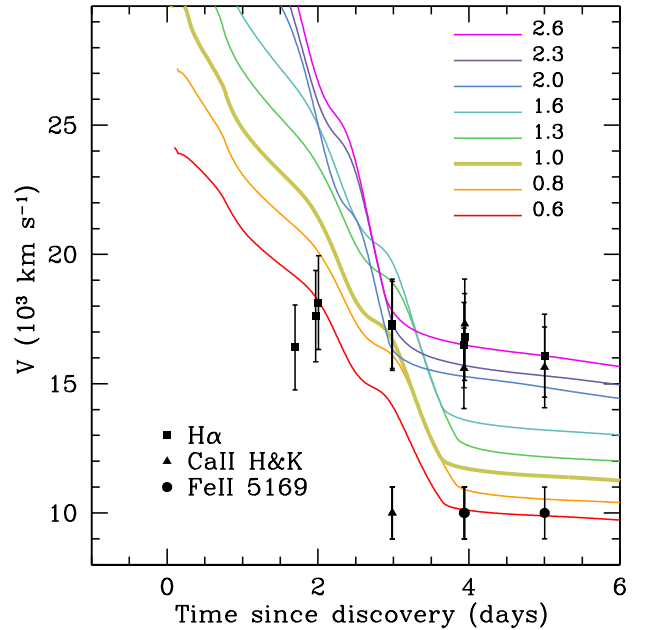}
\caption{The same as Figure \ref{fig:velocities}, but in this case comparing to the best fit wind models at each energy.}
\label{fig:velocities_wind}
\epsscale{1.0}
\end{figure}

We also plot in Figure \ref{fig:velocities} the photospheric velocities inferred from our explosion modeling. These are defined as the velocity at the radius where the optical depth satisfies
\be
	\tau = \int_r^\infty \kappa \rho dr = 2/3,
\ee
where $\kappa$ is the specific opacity. Each color line in \mbox{Figure \ref{fig:velocities}} represents a best fit model to the photometry for a given value of $E_{51}$. The specific value of $E_{51}=1.3$ is represented with a thicker line to highlight that this was our best fit model overall. The numerical models generically show a large gradient in velocity as the photosphere transitions from the low density extended material into the higher density helium core. A comparison between the numerical results and the data show that none of the calculations are consistent with the observations. Even the lowest energy explosion we consider overpredicts the velocities at the earliest times. The best fit explosion energy does even worse. We make a similar comparison in Figure \ref{fig:velocities_wind} for our wind models to emphasize that this problem cannot be reconciled by using a different density profile. At the earliest times, high velocities can cause the the lines to become diluted. This can cause the velocity measurements to be more difficult to make, but it does not appear to explain the discrepancy we find here.

How can these differences between the theory and observations be reconciled when the photometric fits seem reasonable? One resolution would be if the hydrogen-rich material of this SN is not distributed spherically symmetrically. In such a case, the material we are modeling to produce the first peak is indeed moving faster and is hotter than the material represented by the ${\rm H}\alpha$ absorption, but the two components have different angular distributions. This is interesting because it might provide an important clue about the nature of hydrogen-rich material surrounding the helium core, which is related to the progenitor scenarios. For example, one could expect for a wind or strong mass transfer event that the material would be denser (and thus able to sustain the shock generated by the SN) in some regions rather than others depending on the binary configuration.

Alternatively, the asymmetries we are inferring here may point to the generation of deeper asymmetries from the explosion itself. Light echoes from the SN that generated the Cas A remnant show that it was a SN IIb as is being studied here \citep{Krause08,Rest08,Rest11,Finn16}. Detailed studies of the distribution of material and kinematics in the remnant show that the explosion must have been very asymmetric with some indication that it could have even had a jet-like component \citep{Milisavljevic13,Milisavljevic15,Fesen16}.

{Cas A is additionally interesting because the region within its remnant does not show evidence for a massive ($\gtrsim1\,M_\odot$) companion at the moment of explosion \citep{Kochanek17}. This is surprising both because models of SN IIb progenitors typically invoke a binary origin \citep[e.g.,][]{Benvenuto13,Yoon17}, but also because the great majority of massive stars are in binaries \citep{Sana12}. One way to reconcile this is if Cas A was instead the result of a merger \citep[a solution also suggested by][]{Kochanek17}, so that there was a companion, but it was lost before the SN. A merger might also explain the asymmetries that we infer for SN 2016gkg as well as those seen in the remnant of Cas A. In the future, it will be important to continue looking for companions to nearby \mbox{SNe IIb} to get better statistics on how many were in binaries and how many may be due to mergers.}

Finally, we note that \citet{Folatelli14} also point out that some SNe IIb have strangely low velocities and discuss whether this can be produced by asymmetries. In that case though the features in question are ${\rm HeI}\,5876$ and ${\rm HeI}\,7065$, and they are being measured well after the first peak and close to the second peak. Thus that work is probing different material than the very shallow material we are focusing on here. At larger depths, the low velocities are more natural, and may potentially be attributed to the high ionization energy of helium \citep{Piro14c}, but further work should be done to probe how deep asymmetries may be present in SN IIb ejecta.

\section{Conclusions and Discussion}
\label{sec:conclusions}

We have investigated the constraints placed by the well-resolved first light curve peak of SN 2016gkg by comparing these observations to a large grid of numerical envelope models. We considered both convective and wind density profiles to test whether the specific profile assumed leads to any bias in the inferred properties. We find that the first peak is well-described by extended material with a mass of $\approx0.02\,M_\odot$ and radius $\approx180-260\,R_\odot$. Although these values are independent of the  density distribution, we find that the convective profile gives a somewhat better fit to the data in comparison to the wind profile.

As mentioned in Section \ref{sec:convective}, this mass is just refers to the extended material, so that in a realistic stellar model the total hydrogen mass could be larger. {This radius is consistent with pre-explosion imaging with the  {\it Hubble Space Telescope}  by \citet{Kilpatrick17}, who constrained the radius to be \mbox{$\sim35-270\,R_\odot$.} It is somewhat larger than the early temperature modeling by \citet{Tartaglia17}, who use the analytic work of \citet{Rabinak11} to find $\sim48-124\,R_\odot$. The numerical models provide a much more detailed fit to the data (see Figure \ref{fig:photometry}), even for different density profiles, which helps strengthen the argument for the radius we find.} The measured explosion energy is $E_{51}=1.3$, but this is degenerate with the mass of the helium core, as described by Equation (\ref{eq:ee}). We infer from our modeling that the explosion must have occurred within $\approx2-3\,{\rm hrs}$ of the first observed data point, demonstrating that this event was caught very close to the moment of explosion.

SN 2016gkg also has some of the earliest velocity measurements of any SN IIb, potentially providing a unique look into the details of the hydrogen-rich outer material. Comparing our predicted photospheric velocity evolution to these observed velocities show that the models nearly always predict too high of velocities. We suggest that this may be due to asymmetries in the hydrogen-rich outer material, or even the explosion itself. {We also discuss our results in light of the remnant asymmetries and the lack of a companion for \mbox{Cas A,} which may point to a merger origin \citep{Kochanek17}.}

Asymmetry is, of course, a limitation of the one-dimensional modeling performed in this study. Therefore, our modeling may really only represent the densest, optically thick regions of the hydrogen-rich ejecta, with the envelope mass we infer actually being an upper limit if this material is not distributed the same in all directions. It is an open question how the SN shock would propagate in such a geometry because it is possible that the helium-core ejecta may flow more readily to less dense regions, impacting how well the hydrogen-rich material can be thermalized by the explosion.

Early spectropolarimetry data would be helpful for measuring the strength of these asymmetries as well as following how long they last. This would provide some idea about how far the asymmetries extend into the exploding star. Already, using data at radio and X-ray wavelengths as well as late-time spectra, there are indications of a diversity of circumstellar environments around \mbox{SNe IIb} \citep[e.g.,][]{Chevalier10,Maeda15,Kamble16}. These probe much larger radii and less dense material than the work presented here. Nevertheless, looking for correlations between these studies and the mismatch between theoretical and observational velocities as found here may be one way of teasing out the origin of these different populations.

\acknowledgements
We thank Maria Drout, Chris Kochanek, and Ben Shappee for feedback on a previous draft, and Saurabh Jha for helpful discussions. We thank the Summer Undergraduate Research Fellowship (SURF) program at Caltech, which supported the  internship of M.E.M. at the Carnegie Observatories. We also thank Drew Clausen for generating the $15\,M_\odot$ model and associated helium core with \texttt{MESA} that was used in this work. Support for I.A. was provided by NASA through the Einstein Fellowship Program, grant \mbox{PF6-170148.} D.J.S acknowledges support from NSF grant AST-1517649. The computations were performed on the MIES cluster of the Carnegie Observatories, which was made possible by a grant from the Ahmanson Foundation.

\bibliographystyle{apj}

\end{document}